# INVESTIGATION OF FLOW INSIDE AN AXIAL-FLOW PUMP OF "GV – IMP" TYPE

**ANATOLIY A. YEVTUSHENKO[1], ALEXEY N. KOCHEVSKY[1], NATALYA A. FEDOTOVA[1], ALEXANDER Y. SCHELYAEV[2], VLADIMIR N. KONSHIN[2]**

[1] *Department of Applied Fluid Mechanics, Sumy State University, Rimsky-Korsakov str., 2, 40007, Sumy, Ukraine*
*alkochevsky@mail.ru*

[2] *OOO "TESIS", Office 701-703, Unnatov str., 18, 125083, Moscow, Russia*
*asc@tesis.com.ru, volk@tesis.com.ru, http://www.tesis.com.ru*

**Abstract:** The article describes research of fluid flow inside an axial-flow pump that includes guide vanes, impeller and discharge diffuser. Three impellers with different hub ratio were researched. The article presents the performance curves and velocity distributions behind each of the impeller obtained by computational and experimental ways at six different capacities. The velocity distributions behind the detached guide vanes of different hub ratio are also presented. The computational results were obtained using the software tools CFX-BladeGenPlus and CFX-TASCflow. The experimental performance curves were obtained using the standard procedure. The experimental velocity distributions were obtained by probing of the flow. Good correspondence of results, both for performance curves and velocity distributions, was obtained for most of the considered cases.
As it was demonstrated, the performance curves of the pump depend essentially on the impeller hub ratio. Velocity distributions behind the impeller depend strongly on the impeller hub ratio and capacity. Conclusions concerning these dependencies are drawn.
**Keywords:** axial-flow pump, guide vanes, impeller, performance curves, velocity distributions, CFX-TASCflow.

## 1. Introduction

Axial-flow submersible pumps are widely used in Ukraine, in particular, as units of pump stations for land iggiration and drainage. Nowadays, these stations are completed with imported pumps, as Ukrainian enterprises do not produce such pumps. In USSR, such pumps were developed 30 years ago and were produced by the factories "Uralhydromash" (Russia) and "Moldavhydromash" (Moldova). Quality of their design does not correspond to the modern scientific and technical level.

Taking into account large industrial, scientific and technical potential of Ukraine, it would be expedient to organize here own production of such pumps. With this purpose, the research was conducted at the department of applied fluid mechanics, resulted in creation of a perfected design scheme of such pumps named GV – IMP (guide vanes – impeller) [1]. As it was shown [2], the created pumps have lower production costs, require lower maintenance expenses and, at the same time, ensure larger efficiency. And as the pumps of this type are often of large power, efficiency is one of the most important criteria of their perfection.

A new scientific idea implemented in design of these pumps is creation of pre-swirl at the entrance to impeller, opposite to shaft rotation, using guide vanes (these vanes are simultaneously the supports of the engine capsule). As it was found out [3], presence of pre-swirl upstream of the impeller influences significantly the performance curves of the pump, by changing its shape and width of working capacity range.

This article presents the results of further experimental researches aimed at study of influence of geometrical parameters of hydraulic components of the pump upon the flow pattern inside it and its performance curves. This will permit, in particular, to find out minimum hub ratio behind the guide vanes that doesn't cause reverse flow near the hub and find out impeller hub ratio that corresponds to the highest pump efficiency.

For performing the computational research, we have chosen the software package CFX-TASCflow (http://www-waterloo.ansys.com/cfx/), as according to the review [4], this package is one of the most respected CFD tools for simulation of fluid flows in hydraulic machinery. As for the experimental research, performance curves of the pump were obtained using the standard procedure. Besides, we have probed the flow downstream of the impeller with a 5-channel probe at different capacities, thus having obtained experimental distributions of velocities at different modes of pump operation.

## 2. Description of the experimental research

***Object of research.*** For performing the research, a model pump was designed, which hydraulic components are presented at Fig. 1. Those hydraulic components were guide vanes, impeller and discharge conical diffuser.

Guide vanes (Fig. 2) include five cylindrical vanes of 4 mm in thickness, with plane initial section of 15 mm in length. Radius of cylindrical surface used for shaping a vane was 60 mm, spanning angle was 84°, i.e., the inlet edge of each vane was installed streamwise, whereas the outlet edge was almost perpendicular to the flow.

Diameter of each impeller was 180 mm. Each of tested impellers had 4 blades designed according to the method of Voznesensky – Pekin [5] for conditions of axial outflow and constant velocity moment of flow upstream of impeller. Impellers differed by the hub ratio $\bar{d}_{hub}$ – 0.5, 0.39 and 0.28 correspondingly. First two impellers were designed basing upon the meridional projection for the first impeller and, thus, had equal blade profiles. The third impeller was designed basing upon its own meridional projection.

Discharge diffuser was conical in shape and had area ratio of 1.72 and internal angle of 24°. According to the results of the paper [6], the losses in the discharge diffuser of this pump are minimal just with this internal angle.

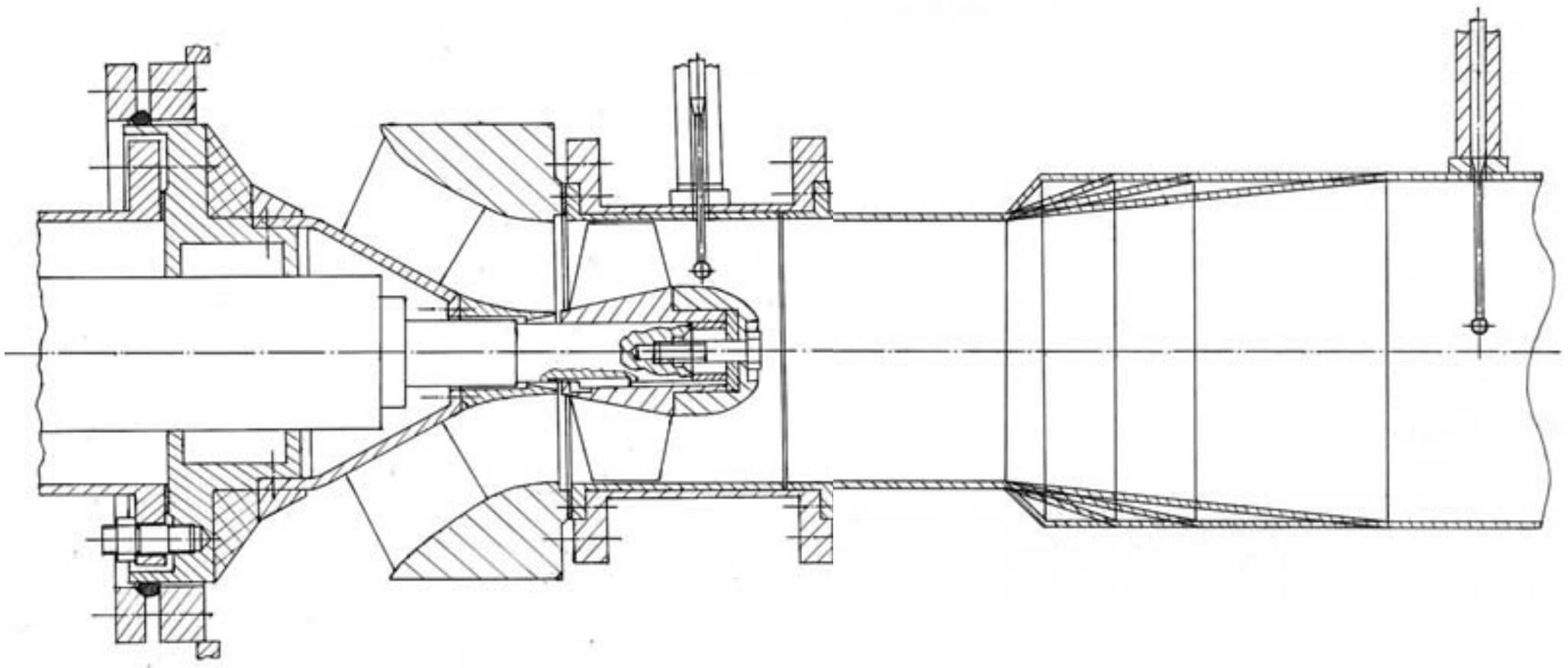

**Figure 1.** Meridional view of the investigated model pump of "GV – IMP" type

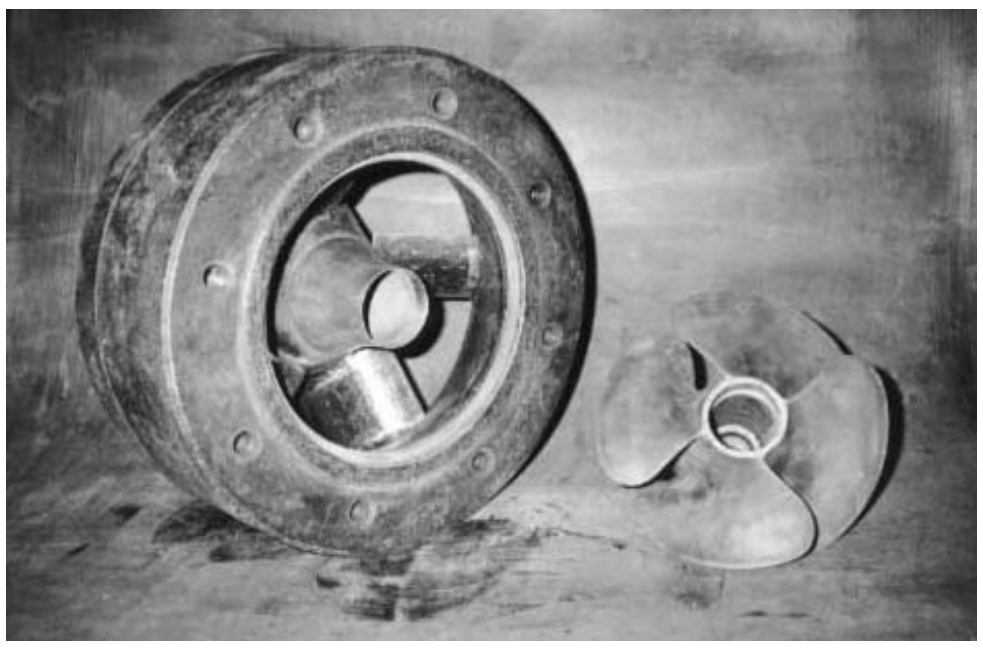
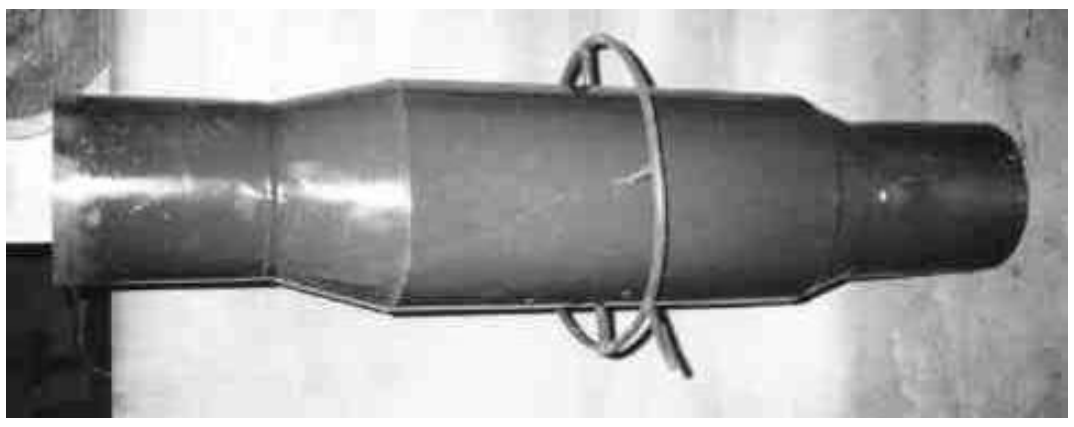

**Figure 2.** Guide vanes, impeller and discharge diffuser (photo)

***Methods of research.*** Performance curves of the pump were obtained by standard test procedure. The flow pattern downstream of impeller was probed using a 5-channel probe. The section of probing was located just behind the impeller blades (Fig. 1). Difference in capacity measured with a diaphragm and computed using the results of probing did not exceed 6%. Description of the probing procedure and formulas for computation of capacity are presented in [7].

## 3. Description of the computational research

For performing the computational research, we have used the software package CFX-TASCflow. The general sequence of actions and separate software tools are described below.

***CFX-BladeGen.*** For creation of solid models of hydraulic components, we have used the software tool CFX-BladeGen. The models of guide vanes and impeller (together with the discharge diffuser) were created separately. The interface of CFX-BladeGen has allowed for input conveniently of all the data from the theoretical drawing of an impeller. Impeller blades were specified by a set of drawing views of profiles. Each profile was obtained at sections of a blade by cylindrical surfaces of different radius. The window of CFX-BladeGen is presented at Fig. 3.

***CFX-BladeGenPlus.*** After creation of solid models of guide vanes and impellers, we have computed flow field inside these components using the software tool CFX-BladeGenPlus. This software tool is featured with easy understandable interface convenient for an engineer without special knowledge in the CFD.

Before computation of flow in CFX-BladeGenPlus, unstructured mesh with tetrahedron cells is generated. We used the meshes containing about 240 000 cells. As source data, fluid properties, capacity, rotational speed and inlet velocity profiles were specified. When computing the flow though guide vanes, we assumed that the upstream flow does not swirl and is of constant velocity though the inlet section. When computing the flow through impellers, we specified at the inlet the velocity distributions obtained by probing the flow behind guide vanes (to be more exact, behind the impeller with the blades removed – these distributions are presented at Fig. 5).

As a result of computation, we obtained the distribution of velocities and pressure through the whole space inside the hydraulic component. We obtained also the integral parameters of flow: axial force and torque imposed on blades, loss factor (for guide vanes), as well as head, consumed power and efficiency (for impellers). The formulas for computing these parameters are editable by a user.

Note, however, that the software tool CFX-BladeGenPlus, being a very convenient tool for express analysis of flow, has restricted possibilities. This tool is designed for computation of flow in detached hydraulic components, not allowing for simulation of flow in the whole flow passage. In this tool, only an algebraic eddy-viscosity turbulence model is implemented, which is too simplified approach for complex flows. Besides, this tool does not allow for simulation of flows with several phases, heat transfer and other effects that require additional model equations. For such problems, CFX-TASCflow should be used.

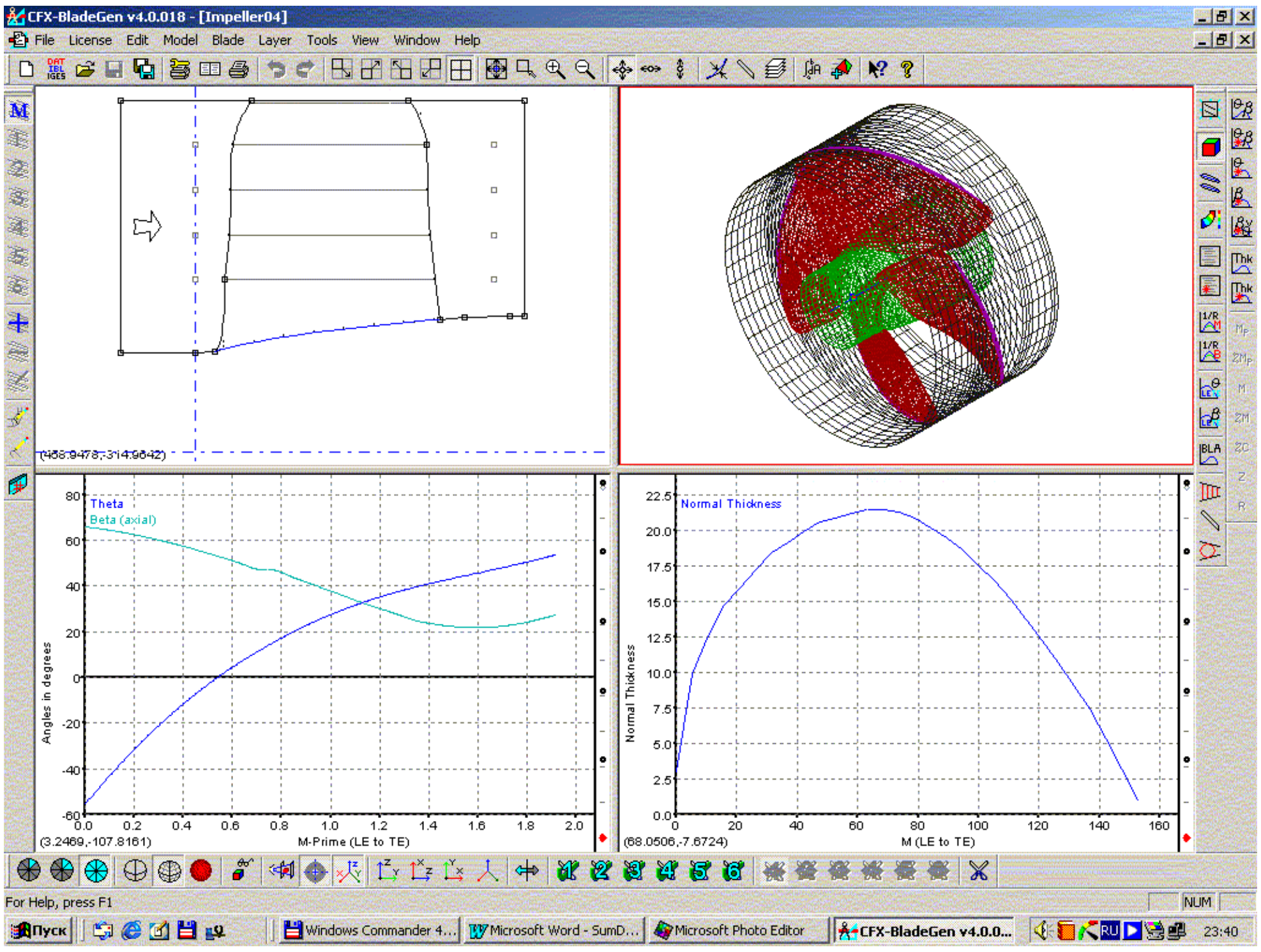

**Figure 3.** Window of the software tool CFX-BladeGen.
In the window, the geometrical model of the impeller with hub ratio of 0.39 is presented.
The dependencies below (blade angles and blade thickness vs. distance along the blade chord) relate to the profile located at the hub.

***CFX-TurboGrid.*** Before computation of flow in CFX-TASCflow, computational mesh should be generated. A convenient tool for generating the mesh in the bladed components is the software tool CFX-TurboGrid. As source data, this tool takes the files created in CFX-BladeGen, and saves the generated mesh in the format required for CFX-TASCflow. CFX-TurboGrid generates structured meshes with hexahedron cells.

The computational domain that corresponds to a separate hydraulic component (guide vanes or impeller) is split into blocks (sub-domains), according to the topology of splitting selected by user. After selection of the topology, the user corrects manually the position of sub-domains, laws of node distribution along grid lines, position of control points, according to the recommendations described in the user manual of CFX-TurboGrid [8]. Each of the topologies implemented in CFX-TurboGrid [8] suits mostly for a certain class of bladed components, providing the possibility to generate high-quality computational mesh with minimal skew of cells.

In this research, we used the following grid topologies: for guide vanes – High Stagger Blade Template, for impellers – Single Block Grid Template. The obtained meshes are presented at the Fig. 4.

Diagnostics of the generated mesh for the guide vanes: total number of cells is 120 000, minimal angle is 20.0°, maximal angle is 163.8°. The mesh for the impeller with $\bar{d}_{hub} = 0.39$: total number of cells is 130 000, minimal angle is 24.2°, maximal angle is 156.4°. Thus, the quality of generated meshes is good enough for running the numerical simulation.

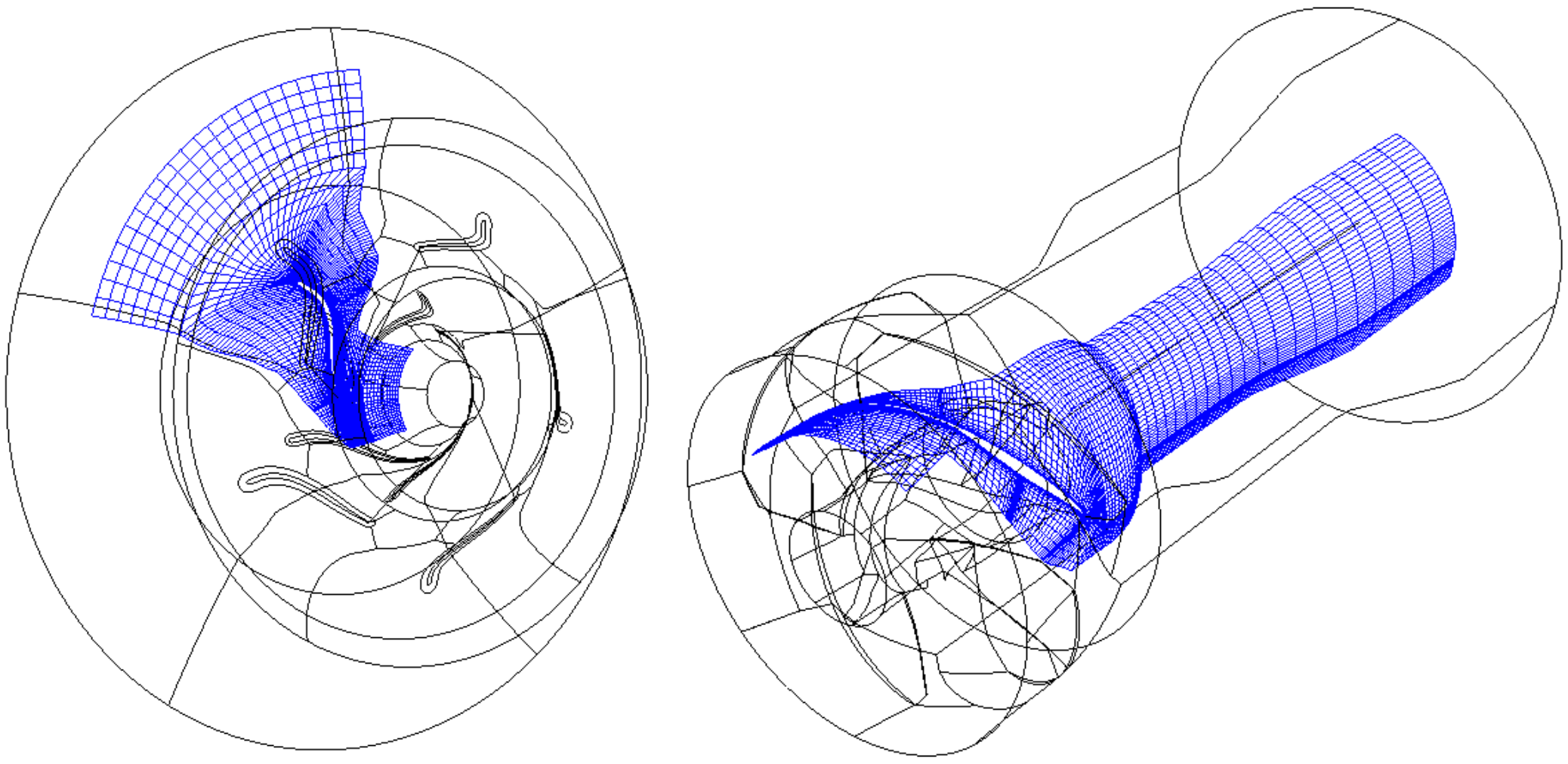

**Figure 4.** Computational mesh for guide vanes (left-hand side) and impeller (right-hand side) – for clearness, only the grid at medium flow surface in one of blade-to-blade channels is shown

***CFX-TASCflow.*** For subsequent actions, the software tool CFX-TASCflow was used.

Firstly, we need to compose the integral computational domain from the separate sub-domains that correspond to guide vanes and impeller (together with the discharge diffuser). The corresponding computational meshes generated by CFX-TurboGrid (Fig. 4) are joined thus forming the united computational mesh. At the interface between guide vanes and impeller, we used the condition of Stage Averaging [4, 9]. Thus, at this surface, the parameters of flow were averaged in the circumferential direction.

In this research, we used $k - \varepsilon$ turbulence model with scalable wall functions. The description of this model and further references are presented, e.g., in [4] and [9].

As the source data for running the simulation, like as in CFX-BladeGenPlus, capacity, rotational speed and fluid properties were specified. The flow upstream of guide vanes was assumed non-swirling, of constant velocity through the inlet section. For turbulence model equations, medium level of turbulence was specified at the inlet (though, variation of this parameter in the wide range almost did not tell upon the results of computation). Zero wall roughness for the whole flow passage was specified. For simplicity, the gap between impeller blades and stator walls was assumed to be zero.

## 4. Results of probing of flow behind the guide vanes

Figure 5 presents comparison of velocity distributions in the section behind guide vanes obtained numerically with CFX-BladeGenPlus and experimentally by probing (the section of probing was the same, only impeller blades were removed). The experimental results of Fig. 5 were earlier published in the paper [10]. Distributions of axial $V_z$ and circumferential $V_u$ velocity are given as ratio to the average (through the section) axial velocity.

As one can see, no separation of flow is observed for all the hub ratios investigated. Distribution of circumferential velocity corresponds approximately to the law of constant velocity moment, $V_u\, r = const$. The numerical and experimental velocity profiles coincide qualitatively, though with some quantitative discrepancy. The swirl intensity obtained by computation exceeds the experimental value.

In order to check grid independence of the solution, this flow was computed using meshes of 120 000, 240 000 and 480 000 cells. The maximum difference in local values of velocities did not exceed 5% from the axial velocity average through the section.

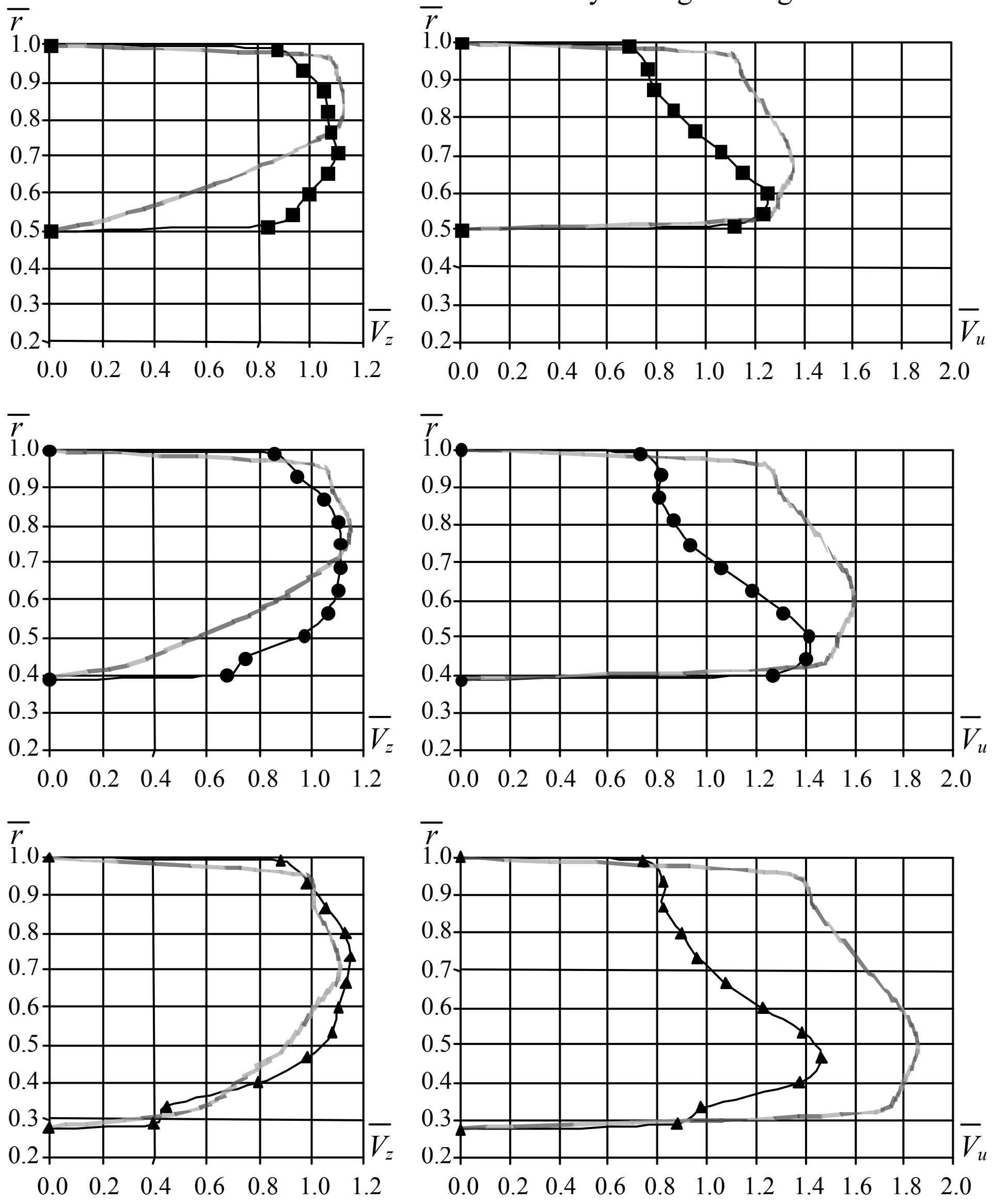

**Figure 5.** Distributions of axial (left-hand side) and circumferential (right-hand side) velocities: experiment: ■ – hub ratio of 0.5, ● – 0.39, ▲ – 0.28 (thin lines are for visual aid only); computation with CFX-BladeGenPlus – pale solid line

## 5. Performance curves of the pump and flow pattern behind the impeller at different capacities

***Dependence of theoretic head of the impeller on capacity.*** As the tool CFX-BladeGenPlus allows for flow simulation only in detached components, we provide comparison with the experimental results by the theoretic head produced by an impeller. Theoretic head is the head divided by hydraulic efficiency.

This dependence of theoretic head on capacity obtained from the numerical and experimental results is presented in the non-dimensional form at Fig. 6. The average divergence of theoretic head obtained with CFX-BladeGenPlus and experimentally for the investigated range of capacity was 10%, the maximal divergence was about 20%. Theoretic head obtained with CFX-TASCflow coincided with the experimental values still better.

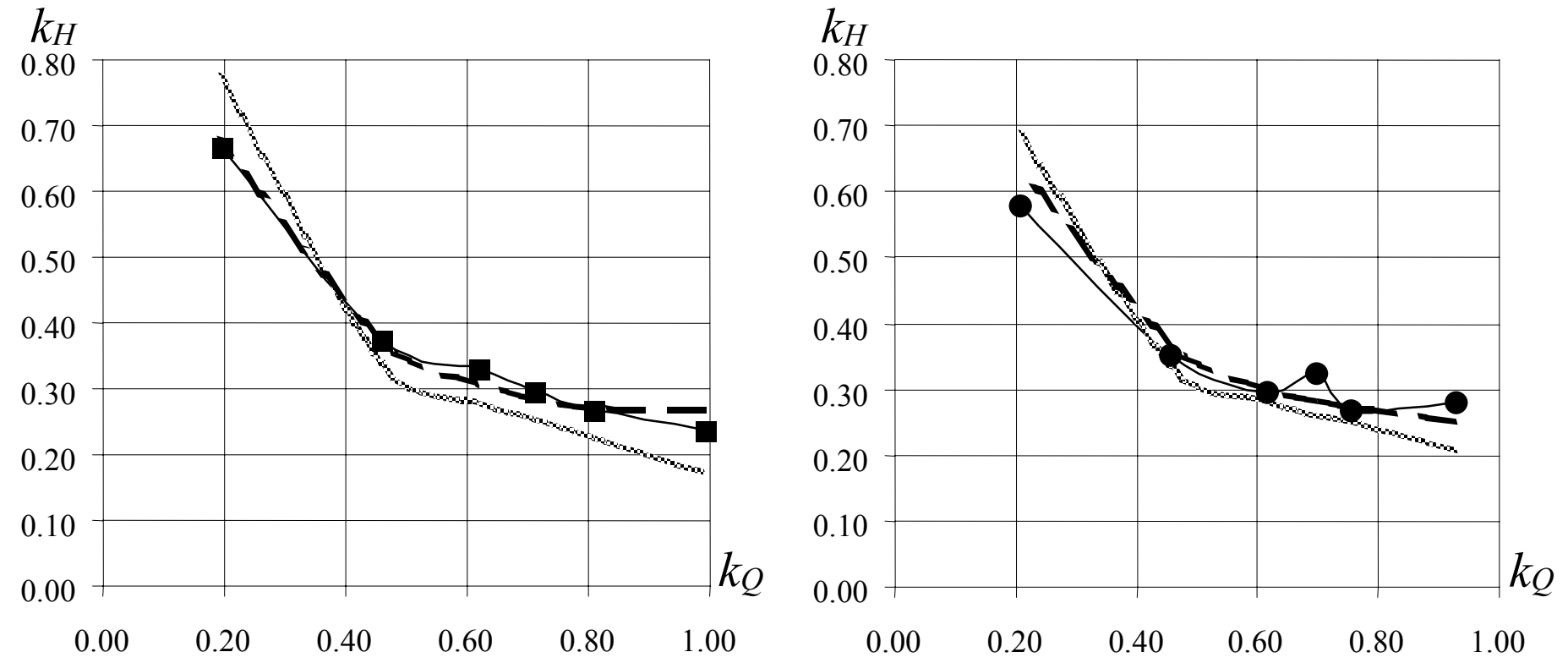

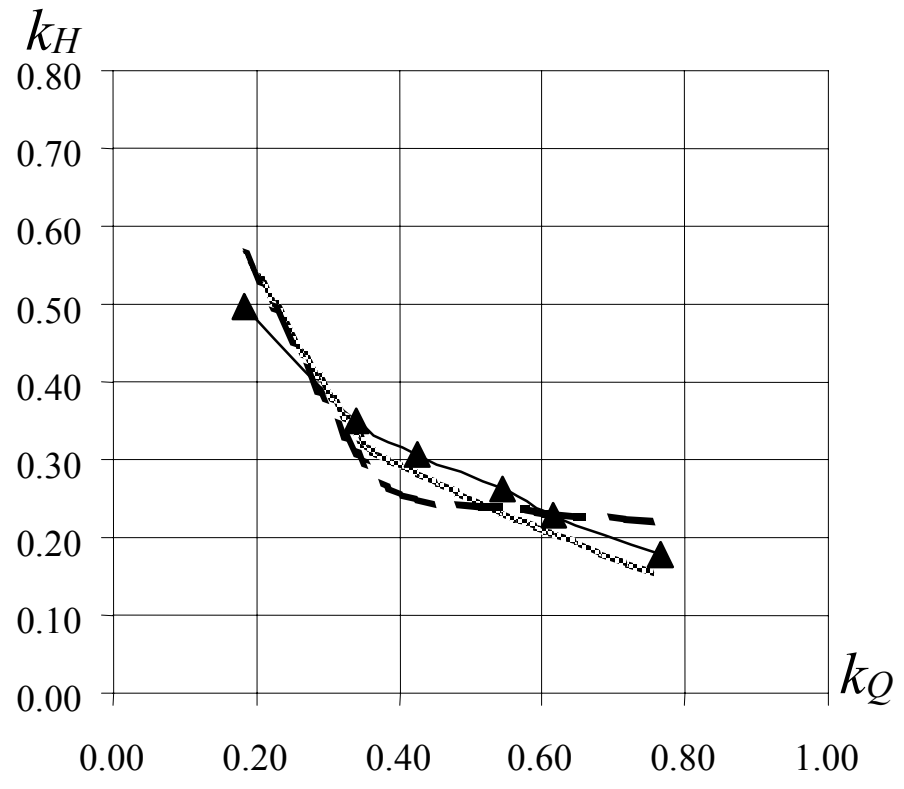

$$K_Q = \frac{Q}{nD^3},$$

$$K_H = \frac{H}{n^2 D^2},$$

$Q$ – capacity, $H$ – head, $n$ – rotational speed, rps, $D$ – impeller diameter

**Figure 6.** Dependence of theoretic head on capacity for the investigated impellers: experiment: ■ – impeller hub ratio of 0.5, ● – 0.39, ▲ – 0.28; computation with CFX-BladeGenPlus – pale solid line; computation with CFX-TASCflow – dashed line

***Performance curves of the pump.*** These curves obtained numerically and experimentally for the whole flow passage of the pump are presented at Fig. 7. Computation of flow using CFX-TASCflow was performed for the same capacities at which probing of flow was performed. As one can see, for the whole investigated range of capacities, the performance curves correspond with each other qualitatively quite well, and in most cases, good quantitative correspondence is also observed.

Fig. 7 presents also the dependence of impeller efficiency on capacity obtained with CFX-BladeGenPlus. The obtained values of efficiency are believable and agree well with the impeller efficiencies obtained using CFX-TASCflow (not shown).

As for the performance curves, the following may be observed. As the capacity of pump increases, the head decreases, and the power increases, as is typical for radial-flow pumps. The larger is the impeller hub ratio, the more extended are performance curves of the pump along capacity axis. The largest efficiency was obtained for the impeller hub ratio of 0.39, however, in this case, capacity range with high level of efficiency was the narrowest. The largest consumed power was also obtained for the impeller hub ratio of 0.39.

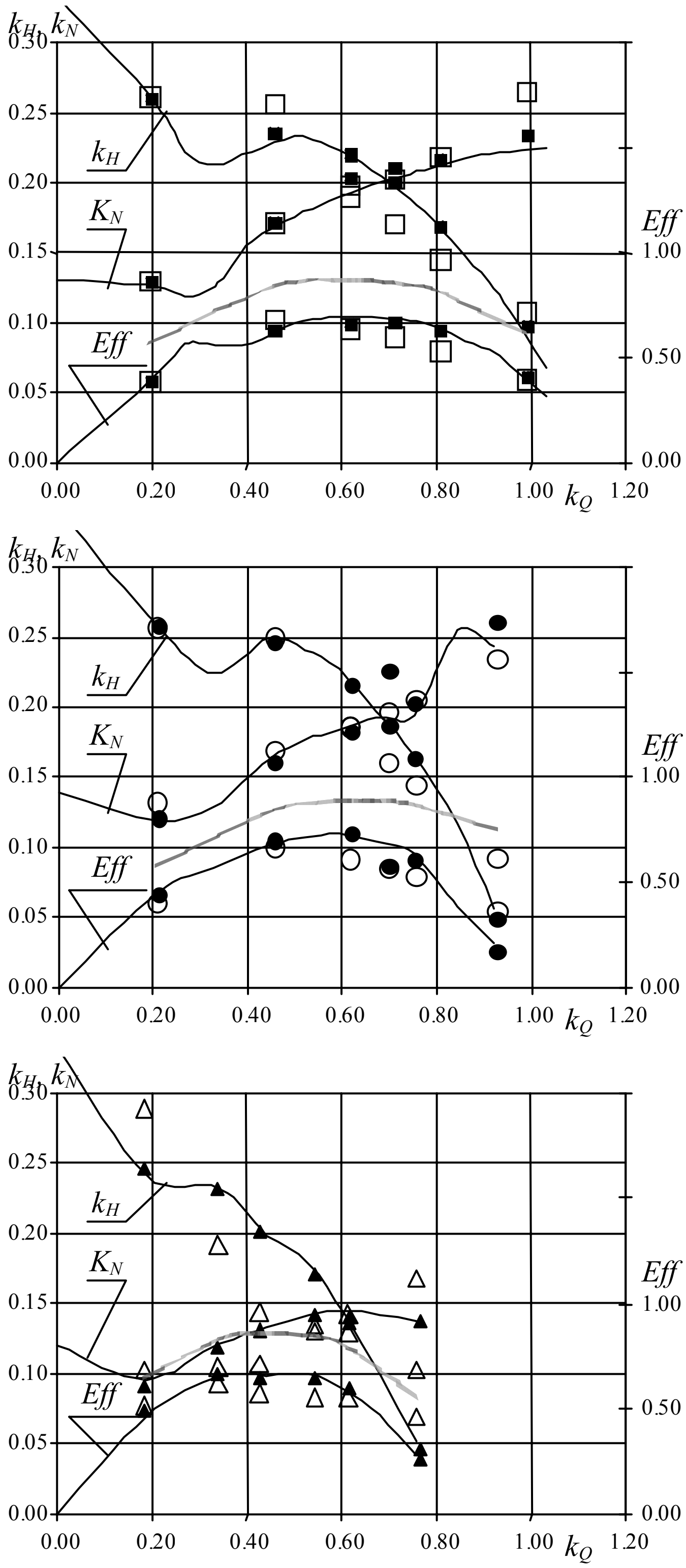

Impeller with the hub ratio of 0.5, the modes for which probing was performed:

1 – $K_Q = 0.99$;
2 – $K_Q = 0.81$;
3 – $K_Q = 0.71$;
4 – $K_Q = 0.62$;
5 – $K_Q = 0.46$;
6 – $K_Q = 0.19$.

Impeller with the hub ratio of 0.39, the modes for which probing was performed:

1 – $K_Q = 0.93$;
2 – $K_Q = 0.76$;
3 – $K_Q = 0.70$;
4 – $K_Q = 0.62$;
5 – $K_Q = 0.46$;
6 – $K_Q = 0.21$.

Impeller with the hub ratio of 0.28, the modes for which probing was performed:

1 – $K_Q = 0.76$;
2 – $K_Q = 0.62$;
3 – $K_Q = 0.54$;
4 – $K_Q = 0.43$;
5 – $K_Q = 0.34$;
6 – $K_Q = 0.18$.

**Figure 7.** Dependence of head, power and efficiency on capacity of the pump:
solid lines – experiment – performance curves obtained using the standard technique;
painted markers – experiment – the modes for which probing was performed:
■ – impeller hub ratio of 0.5, ● – 0.39, ▲ – 0.28; blank markers – computation, CFX-TASCflow;
pale line – impeller efficiency, computation, CFX-BladeGenPlus

Maximum efficiency achieved with each of these impellers, according to the experimental results, is 70%, 73% and 67%, and according to the computational results, is 68%, 67% and 62% correspondingly. According to the experiment, maximum efficiency for the impeller with $\bar{d}_{hub}$ = 0.5 is reached at the mode 3, for the impeller with $\bar{d}_{hub}$ = 0.39 – at the mode 4 and for the impeller with $\bar{d}_{hub}$ = 0.28 – at the mode 5. According to the computation, maximum efficiency for each impeller is reached at the mode 5.

Bladed components of this pump are designed in such a way that behind the impeller it would be no swirl at the nominal capacity. While passing between guide vanes, the flow obtains large negative swirl, of approximately constant velocity moment through the section. As the flow passes though impeller, its velocity moment changes to zero. As a result of this, the head is created.

At the capacities above nominal, the flow after passing though the impeller keeps its negative swirl and, thus, it swirls downstream in the direction opposite to shaft rotation. At the capacities below nominal, the flow after passing though the impeller gets positive swirl and, thus, it swirls downstream in the same direction as the shaft. Corresponding flow patterns in absolute frame of reference obtained with CFX-TASCflow are shown at Fig. 8. The more the capacity differs from the nominal, the larger is swirl in the discharge diffuser.

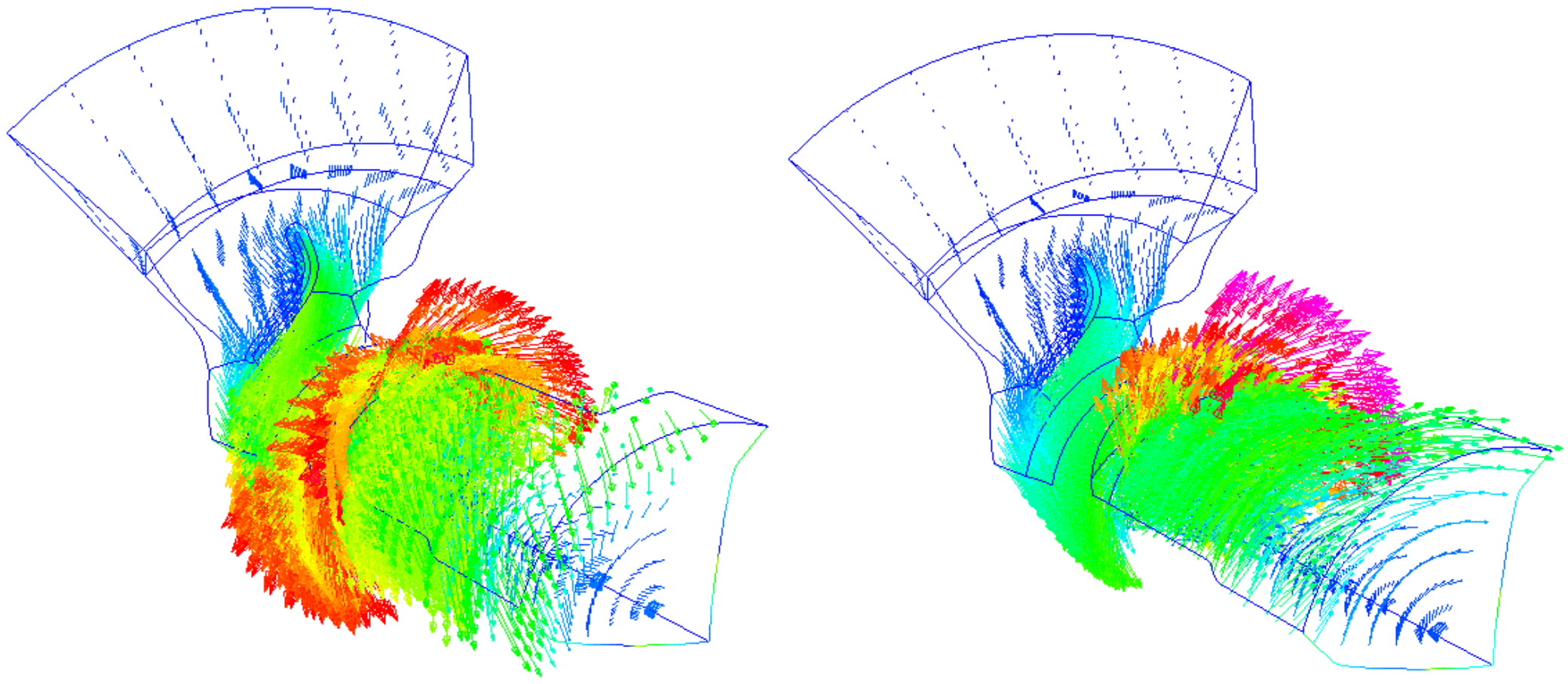

**Figure 8.** Velocity vectors of the flow inside the pump at the modes 1 (left-hand side) and 5 (right-hand side)

***Velocity distributions behind the impeller.*** These distributions obtained by computation using CFX-TASCflow and by probing of flow are presented at Fig. 9. Distributions of axial $V_z$ and circumferential $V_u$ velocity for each impeller are given as ratio to the average (through the section) axial velocity at the nominal capacity for the same impeller. Non-dimensional circumferential velocity at the hub rotating together with impeller, for the mentioned impellers, is 1.31, 1.21 and 1.12 correspondingly.

Experimental results of Fig. 9 at the modes 2, 3 and 4 were earlier published in [10]. As may be seen, the results obtained with CFX-TASCflow, in general, coincide with the experimental results quite well, with proper reflection of changes in velocity distributions at different capacities. The quantitative correspondence of results is also rather good. Significant discrepancies are observed mostly for the impeller with $\bar{d}_{hub}$ = 0.28 and for very low capacities, as strong swirl present in the flow passage in these cases is difficult to be simulated properly and requires advanced turbulence modeling.

The following effects are peculiar to these distributions.

Mode 1. Large residual swirl created by guide vanes is available. The flow is pressed to the periphery, especially distinctly behind the impeller with $\bar{d}_{hub}$ = 0.28, where large stagnation region is observed near the hub.

**Mode 1** (the largest capacity, the head is close to zero):

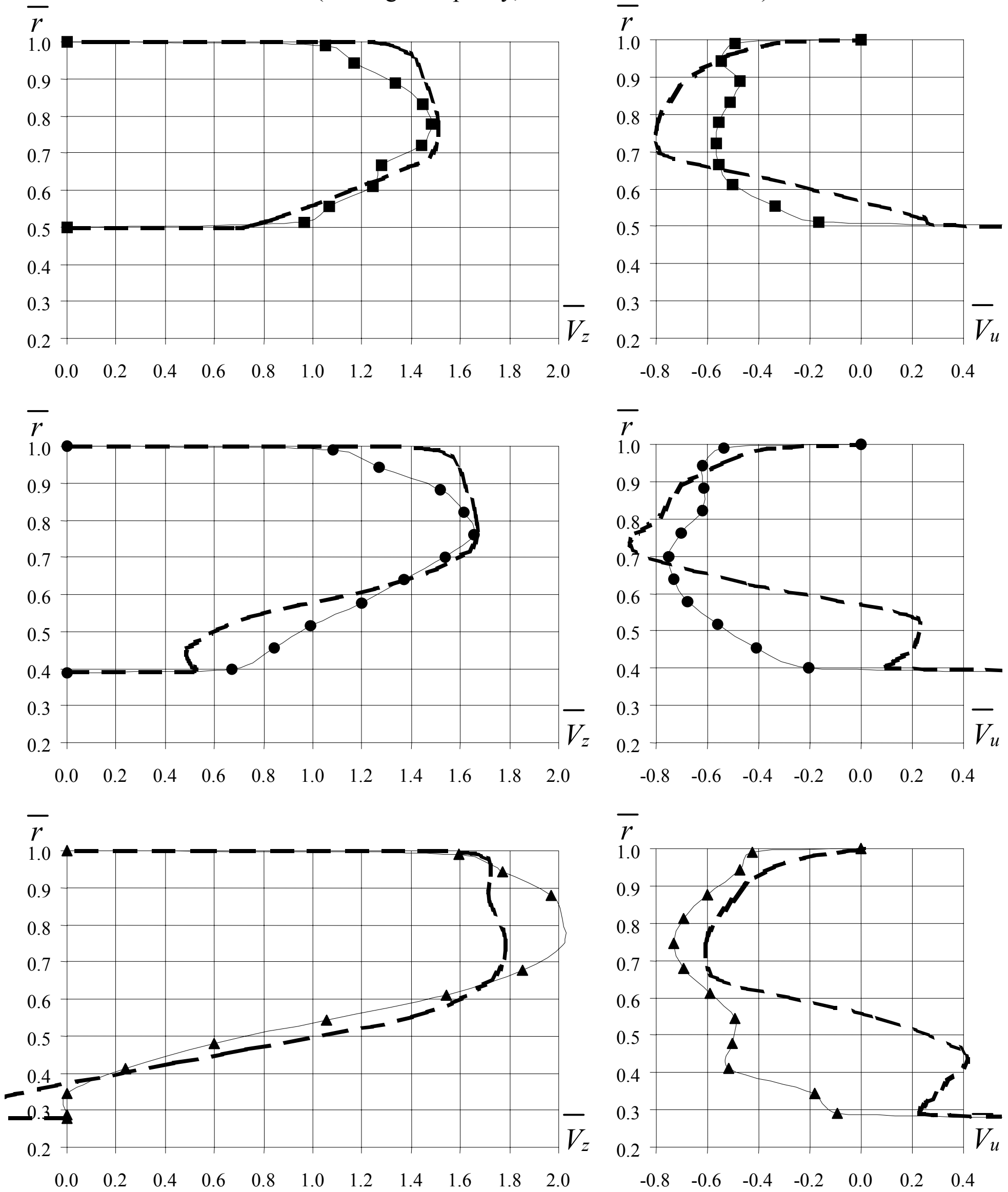

**Mode 2** (the right bound of the working range):

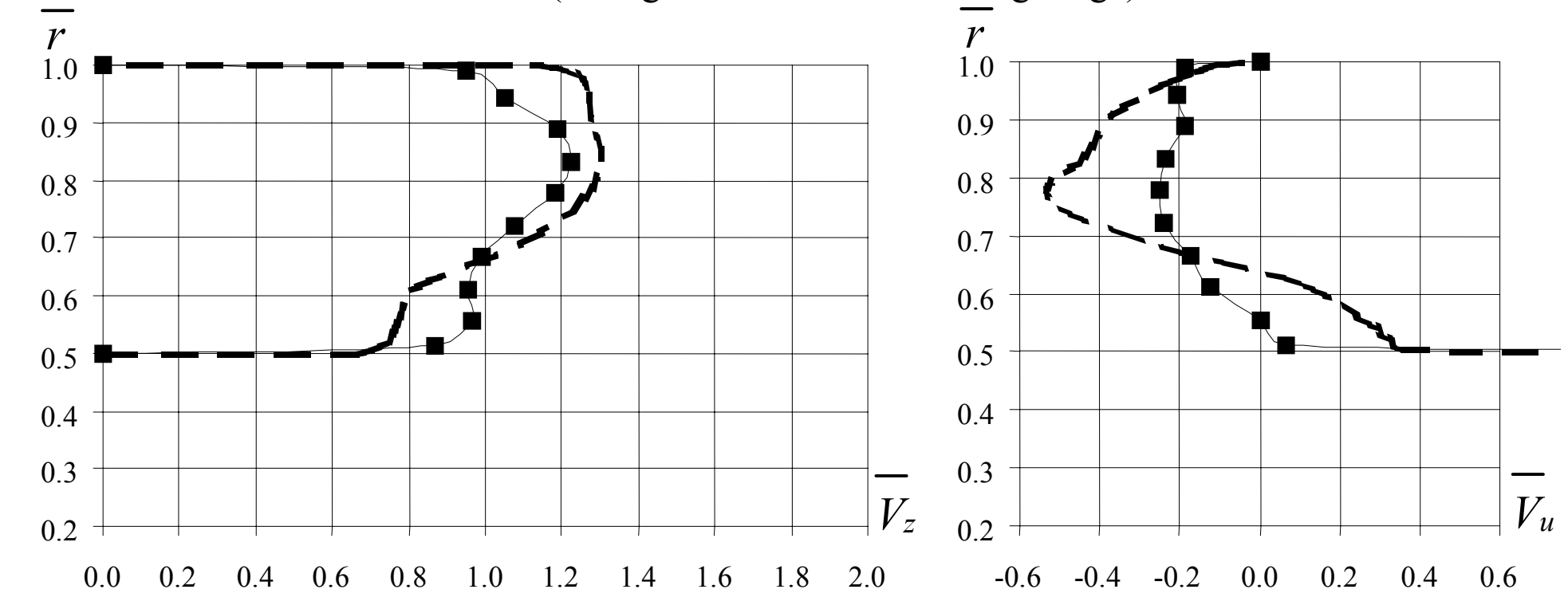

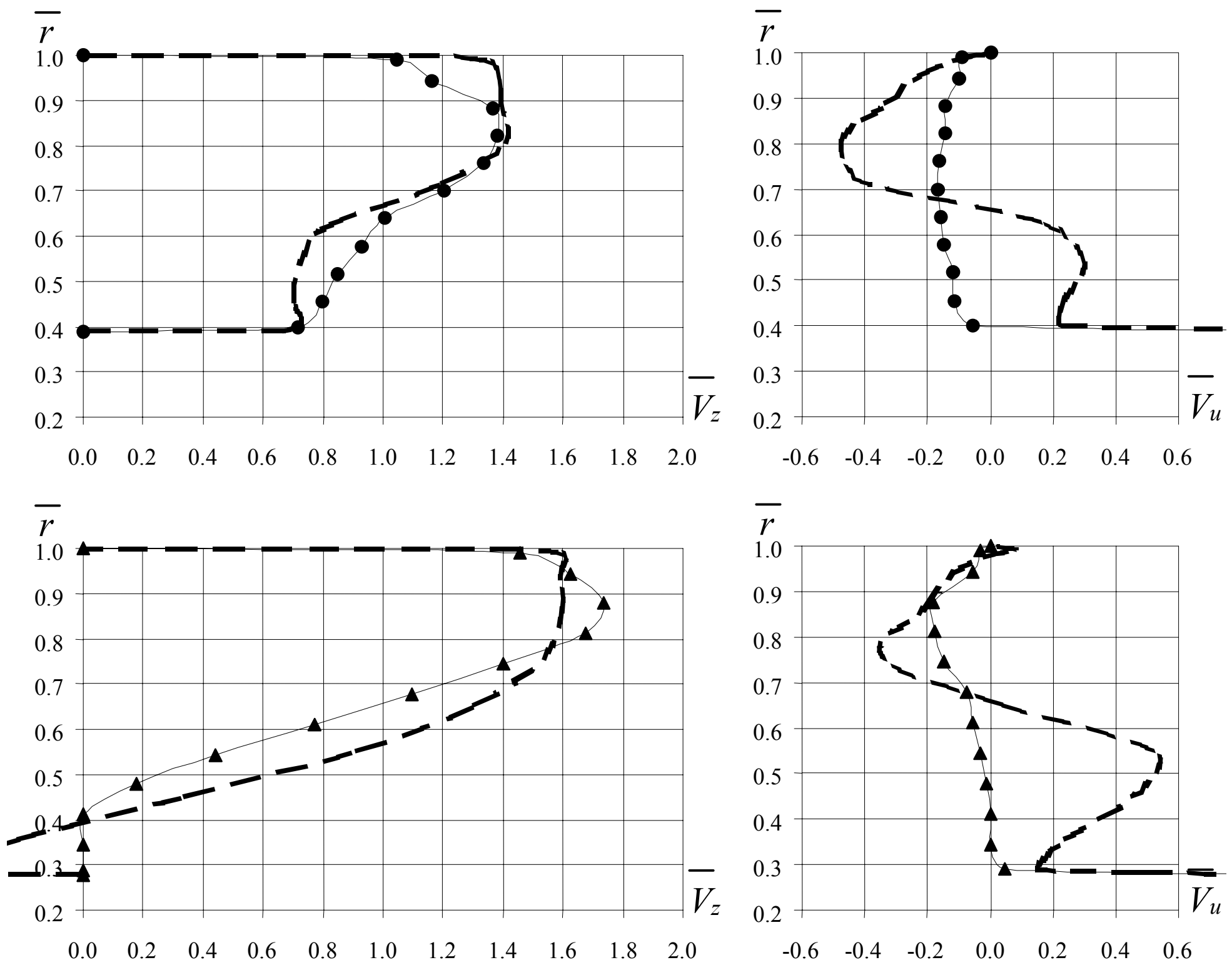

**Mode 3** (the middle of the working range):

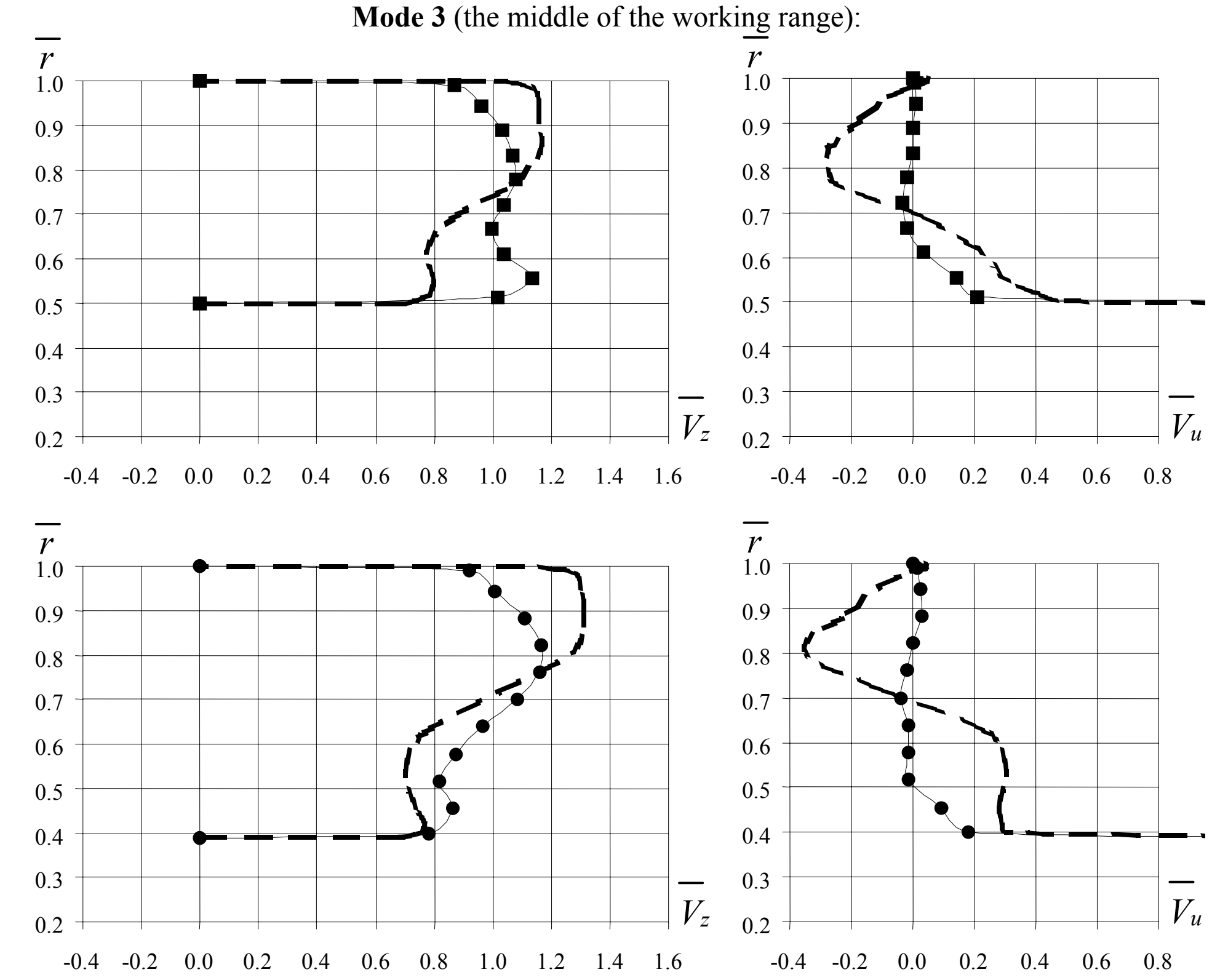

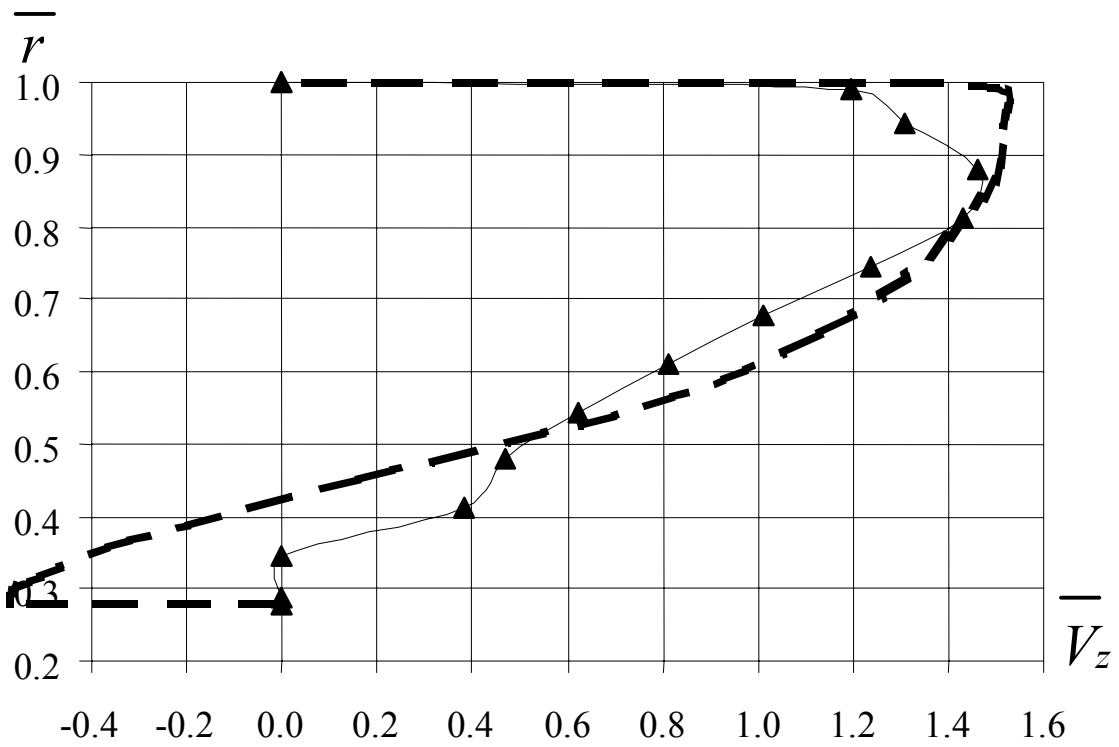

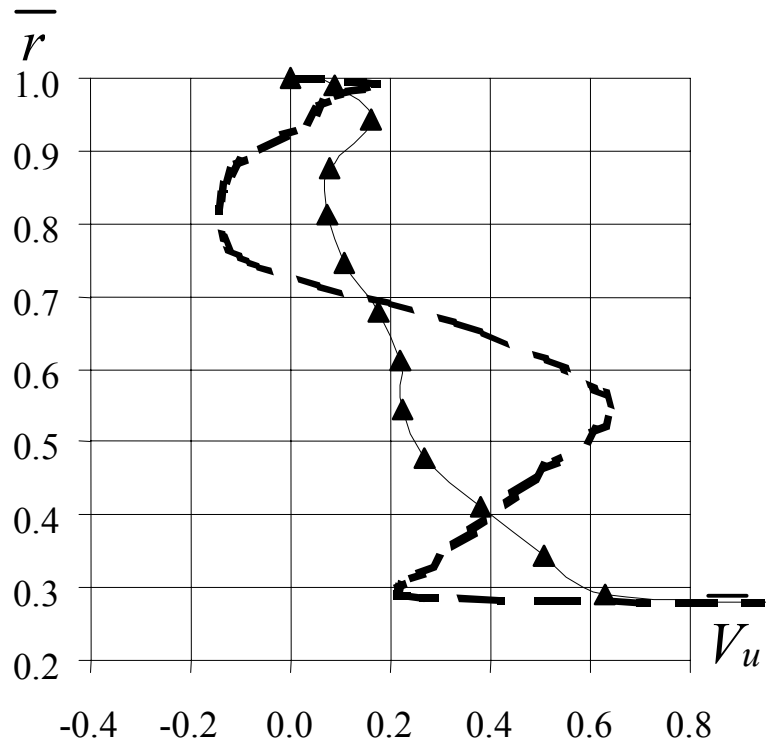

**Mode 4** (the left bound of the working range):

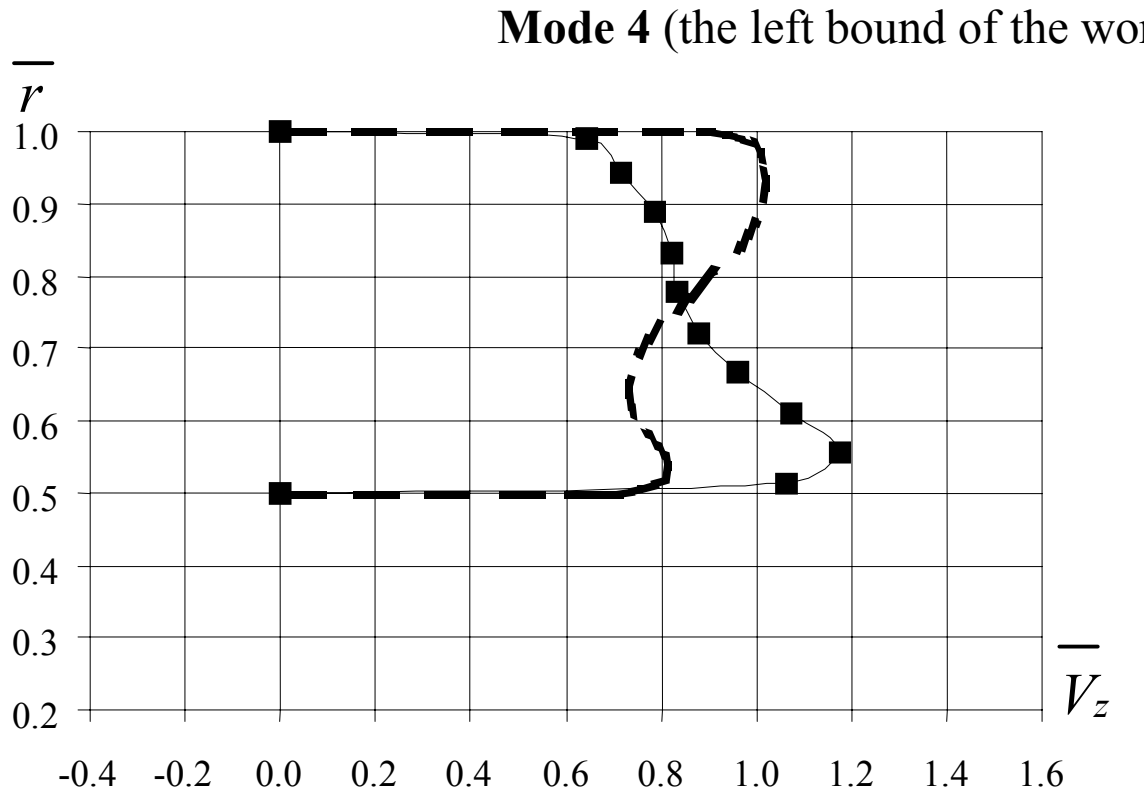

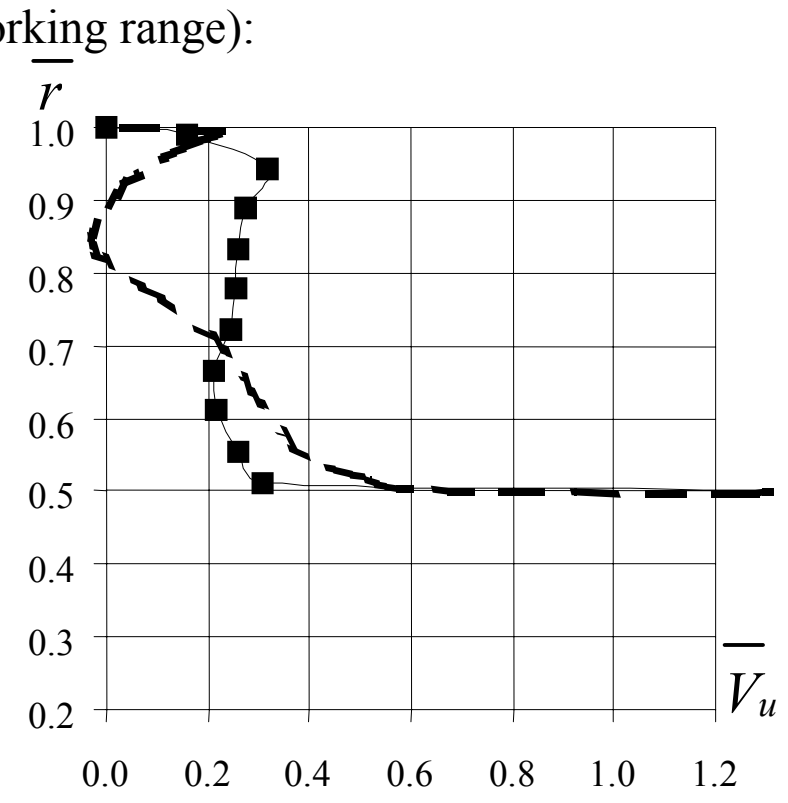

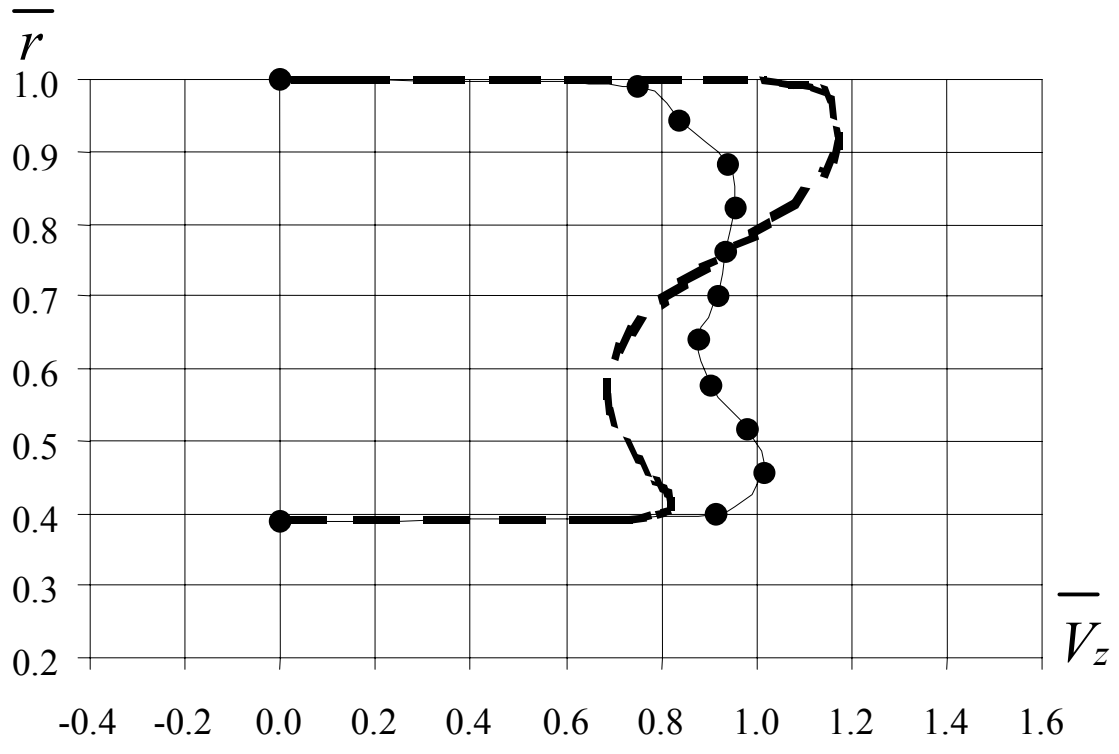

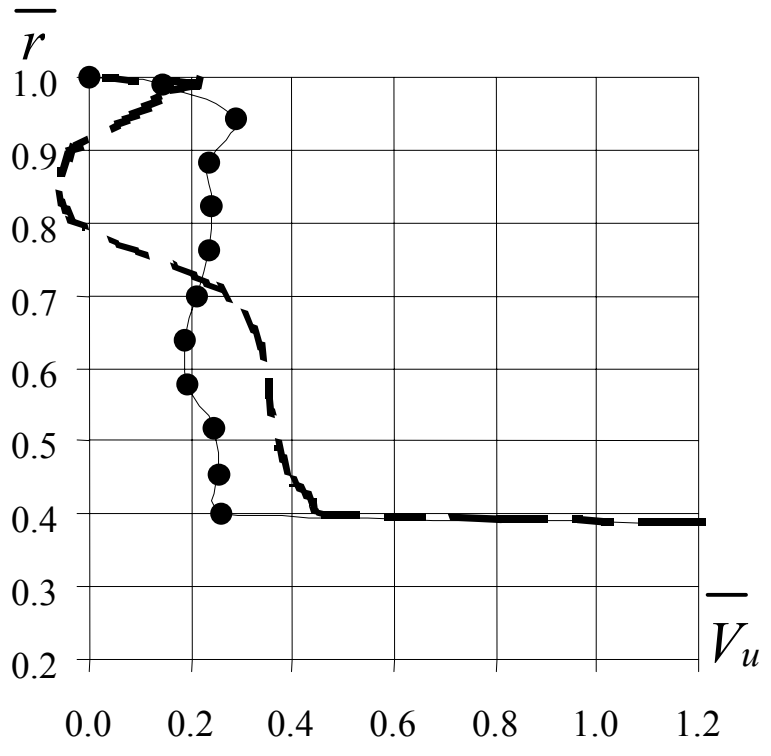

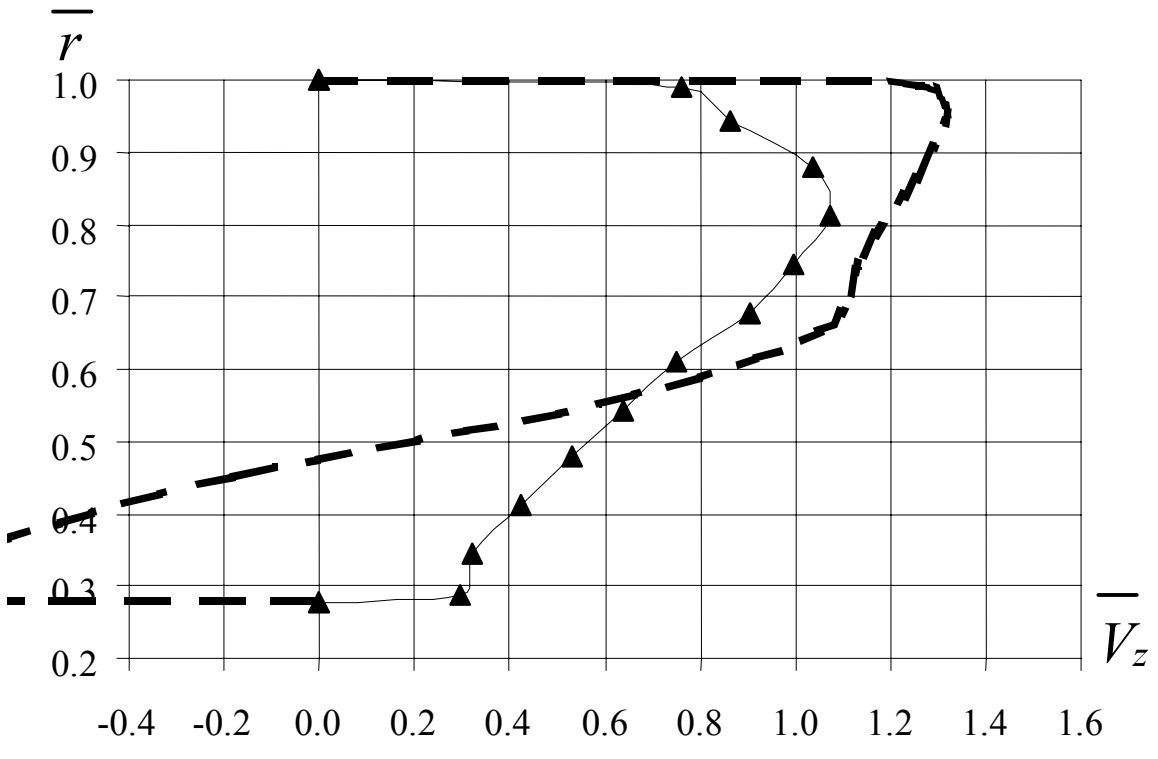

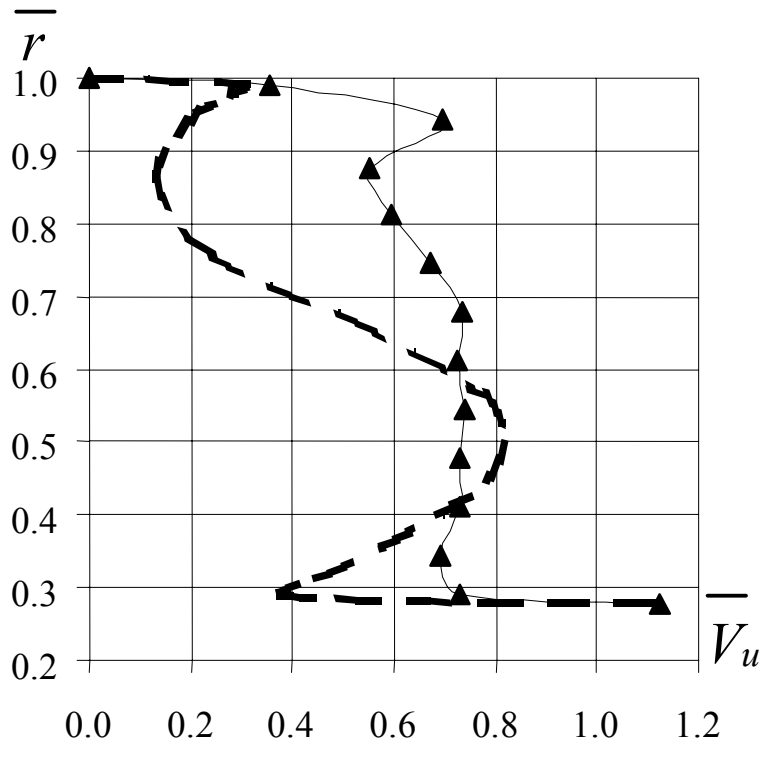

**Mode 5** (low capacity, the mode to the right-hand side from the “pit”):

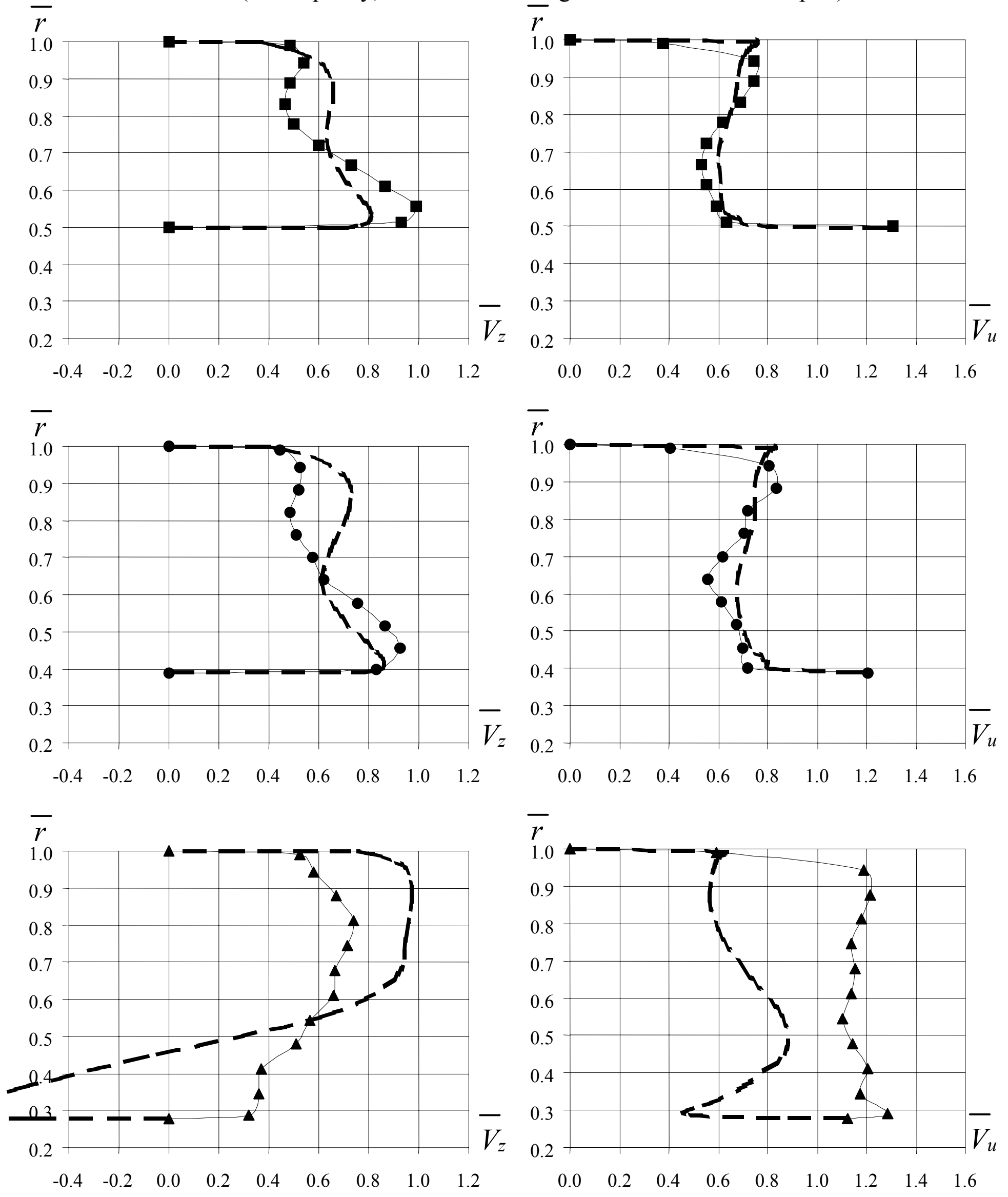

**Mode 6** (the lowest capacity, the mode to the left-hand side from the “pit”):

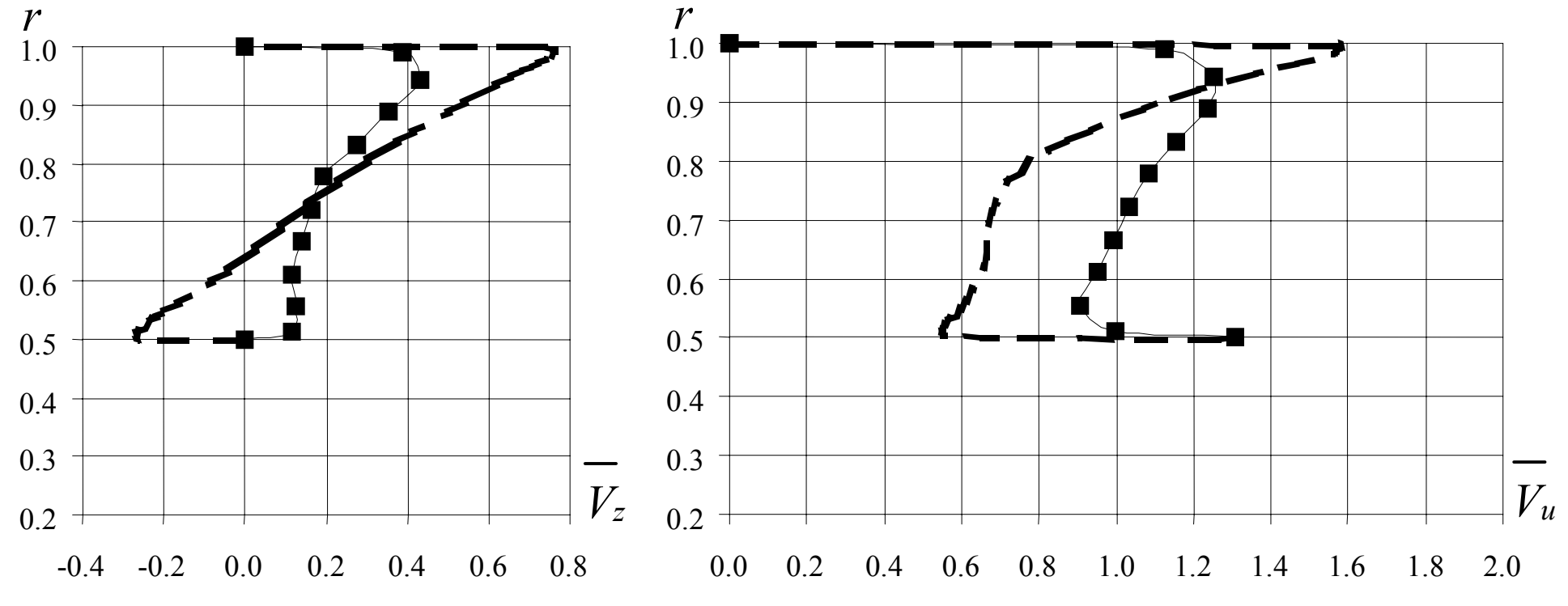

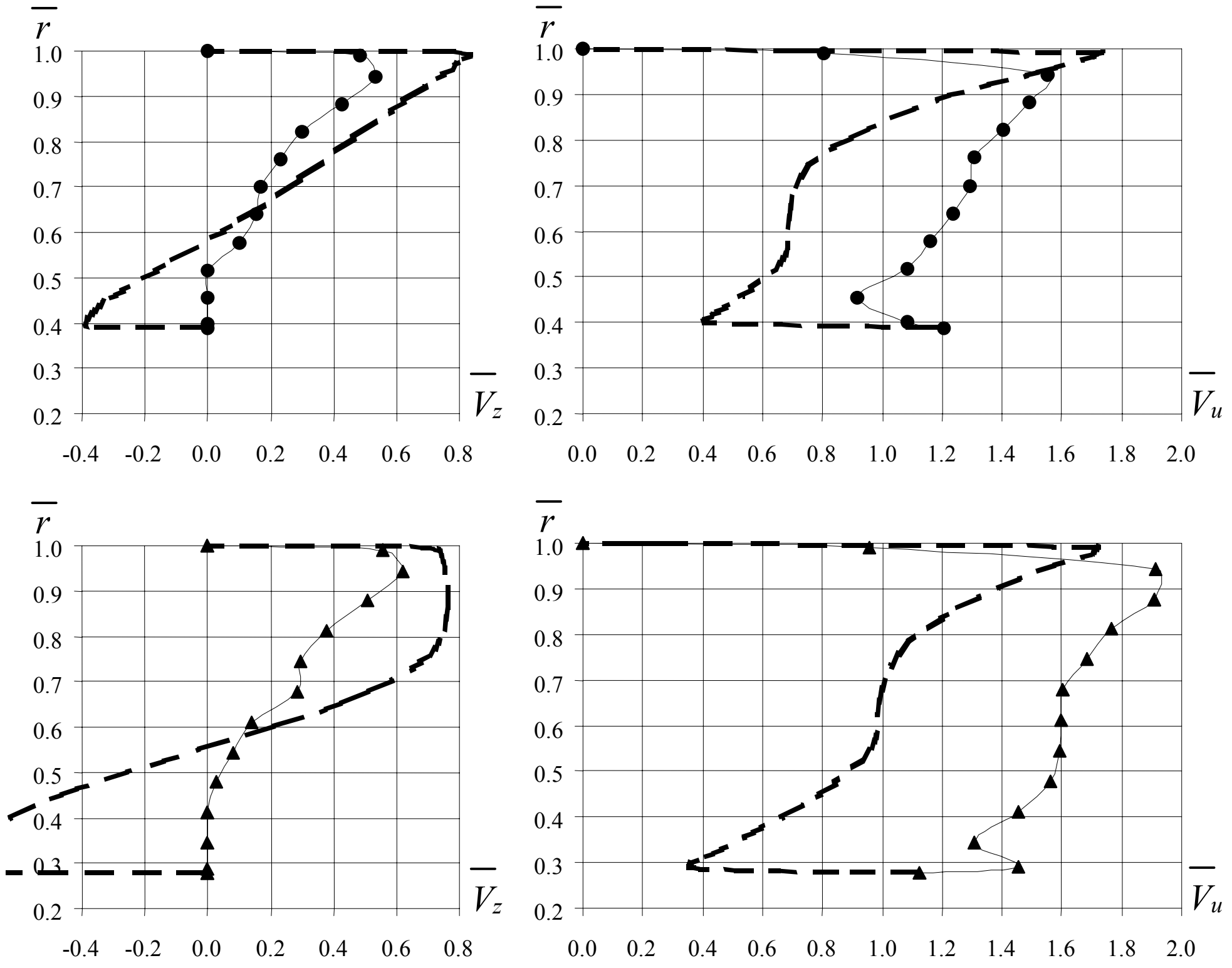

**Figure 9.** Distributions of axial (left-hand side) and circumferential (right-hand side) velocity component behind the impeller: experiment – ■ – impeller hub ratio of 0.5, ● – 0.39, ▲ – 0.28 (thin lines are for visual aid only); dashed line – computation, CFX-TASCflow

Mode 2. At this mode, the flow swirled by guide vanes is also not completely de-swirled after passing the impeller. For each of impellers, the circumferential velocity $V_u < 0$. Axial velocity distribution behind each of three impellers is distinctly pressed to the periphery. Behind the impeller with $\bar{d}_{hub} = 0.5$, it is the most uniform through the section, and behind the impeller with $\bar{d}_{hub} = 0.28$, it is the most deformed.

Mode 3. At this mode, the swirl of flow behind the impeller is the closest to zero. Peak of the axial velocity observed at the previous mode here is less distinct. On the other side, peak of the axial velocity near the hub is observed. It is most distinct behind the impeller with $\bar{d}_{hub} = 0.5$ and merely observable (according to the experimental results) behind the impeller with $\bar{d}_{hub} = 0.28$.

Mode 4. At this mode, the flow swirled by guide vanes is re-swirled after passing the impeller in the direction of shaft rotation. For each of three impellers, the circumferential velocity $V_u > 0$. Peak of axial velocity at the periphery is completely smoothed away at this mode behind the impeller with $\bar{d}_{hub} = 0.5$ (according to the experimental results) but is still distinctly expressed behind the impeller with $\bar{d}_{hub} = 0.28$. The flow behind the impeller with $\bar{d}_{hub} = 0.5$ is strongly pressed to the hub. The stagnation zone behind the impeller with $\bar{d}_{hub} = 0.28$ that was observed near the hub at the previous modes, now is absent (though this fact is not confirmed by computation). Behind the impeller with $\bar{d}_{hub} = 0.39$, peaks of axial velocity at the periphery and near the hub are approximately equal (according to the computation, tendency for alignment of peaks is observed).

Mode 5. Large swirl of flow in the direction of shaft rotation is available. Circumferential velocity is almost constant through the section. In the axial velocity distribution, a sharply expressed peak near the hub is available (except for the impeller with $\bar{d}_{hub} = 0.28$).

Mode 6. In comparison with the previous mode, the intensity of flow swirl has strengthened. Distributions of both axial and circumferential velocities have essentially changed and pressed to the periphery. Flow pattern has become similar to the solid body type of rotation.

Note, at the head curve of the pumps (Fig. 6), between the modes 5 and 6, a "pit" is available (capacity range featuring with reduced head and unstable flow inside the pump). Presence of the "pit" can seemingly be explained by restructuring of flow inside the pump. At the capacities to the right from the "pit", the axial velocity distribution is pressed to the hub, so as at the capacities to the left from the "pit" it is pressed to the periphery. Further downstream, the axial velocity distribution is gradually smoothed approaching finally logarithmic shape typical for developed turbulent flow in a pipe. Process of smoothing of the velocity distribution may be imagined as superposition of flow of constant (through the section) velocity and vortex flow which decelerates rapidly moving fluid layers and accelerates slowly moving fluid layers (Fig. 10). In this representation, when the mode of pump operation changes from 5 to 6, this vortex (that smoothes the axial velocity distribution downstream of impeller) changes the direction of rotation (considering the meridional projection of the pump). Here is the explanation of instability of pump operation at the capacities within the "pit": this vortex rotates sometimes in one direction, sometimes in the opposite direction.

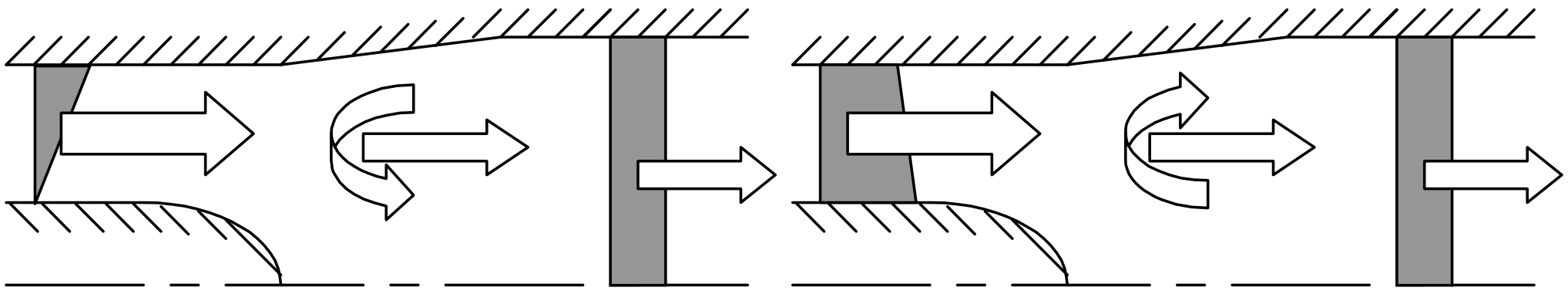

**Figure 10.** Scheme of flow behind the impeller at the mode 6 (left-hand side – to the left from the "pit") and 5 (right-hand side – to the right from the "pit"); the dark figures represent distributions of the axial velocity

In the impeller with $\bar{d}_{hub}$ = 0.28, the fluid flow is pressed to the periphery in the whole capacity range (Fig. 9). Correspondingly, at the head curve of the pump with this impeller, there is no "pit" (Fig. 7).

## 6. Conclusions

As a result of this research, good correspondence of computational results obtained using the software package CFX with experimental results was observed, except for strongly swirled flows. Namely, with CFX-BladeGenPlus the performance curves of the pump and velocity distribution behind the guide vanes were agreed, with CFX-TASCflow – the performance curves and velocity distribution behind the impellers.

The following conclusions can be also drawn:

- When hub ratio behind the guide vanes of this design is 0.28, the reverse flow behind the guide vanes is still absent.
- Flow swirl behind the impeller depends strongly on the capacity of pump. At capacities above nominal, the residual swirl generated by guide vanes is available behind the impeller. At low capacities, the impeller forces the flow to swirl in the direction of shaft rotation.
- Distribution of the axial velocity behind the impeller also depends significantly on the capacity. At high capacities, fluid flow is pressed to the periphery. As the capacity decreases till the mode to the right from the "pit" on the head curve, the flow is gradually depressed from the periphery and pressed to the hub. At very low capacities (to the left from the "pit"), the flow is strongly pressed to the periphery.

- Distribution of the axial velocity behind the impeller also depends on the impeller hub ratio. At the hub ratio of 0.28, the flow is pressed to the periphery in the whole range of capacities.
- Shape of performance curves obtained in this axial-flow pump with large negative inlet swirl is typical for radial-flow pumps (as the capacity increases, the head decreases and the power increases).
- As the impeller hub ratio increases, its performance curves extend along the capacity axis.
- The highest efficiency (experiment – 73%, computation – 68%) was reached at the impeller hub ratio of 0.39 (experiment; computation – 0.5). However, in this case, the range of capacities of high efficiency was the narrowest.
- The largest consumed power was also observed at the impeller hub ratio of 0.39 (experiment; computation – 0.5).
- In the pumps with impeller hub ratios of 0.5 and 0.39, a distinctly expressed "pit" was observed at performance curves. At the impeller hub ratio of 0.28, this "pit" was absent.
- According to the experiment, in the pump with the impeller hub ratio of 0.5, the highest efficiency was reached at approximately zero swirl behind the impeller. At the hub ratio of 0.28, the highest efficiency was obtained at large positive swirl behind the impeller, i.e., at a significantly lower capacity. According to the results of computation with CFX-TASCflow, for each of impellers the highest efficiency was reached at rather large positive swirl behind the impeller.

## *Acknowledgements*

The present research was conducted under support of the collective of the department of fluid mechanics of Sumy State University.